# Human Computer Interaction Design for Mobile Devices Based on a Smart Healthcare Architecture

Authors

Pu Liu [1,2]; ORCID: 0000-0002-8876-0930

Sidney Fels [3]; ORCID: 0000-0001-9279-9021

Nicholas West [1]; ORCID: 0000-0001-8382-5136

Matthias Görges [1,2]; ORCID: 0000-0003-2193-178X

Affiliations

[1] Department of Anesthesiology, Pharmacology & Therapeutics, The University of British Columba, BC, Canada

[2] Research Institute, BC Children's Hospital, BC, Canada

[3] Department of Electrical and Computer Engineering, The University of British Columba, BC, Canada

Corresponding author

Dr Matthias Görges, BC Children's Hospital Research Institute, V3-324, 950 West 28th Ave, Vancouver, BC, V5Z 4H4, Canada; 1-604-875-2000 x5616; mgorges@bcchr.ca

Conflict of Interest: The authors declare that they have no conflict of interest.

Funding: This work was supported, in part, by a Canadian Institutes of Health Research grant (PJT-149042).

Ethical approval: This article does not contain any studies with human participants or animals performed by any of the authors.

# Human Computer Interaction Design for Mobile Devices Based on a Smart Healthcare Architecture

**Abstract**

Smart and IoT-enabled mobile devices have the potential to enhance healthcare services for both patients and healthcare providers. Human computer interaction design is key to realizing a useful and usable connection between the users and these smart healthcare technologies. Appropriate design of such devices enhances the usability, improves effective operation in an integrated healthcare system, and facilitates the collaboration and information sharing between patients, healthcare providers, and institutions. In this paper, the concept of smart healthcare is introduced, including its four-layer information architecture of sensing, communication, data integration, and application. Human Computer Interaction design principles for smart healthcare mobile devices are outlined, based on user-centered design. These include: ensuring safety, providing error-resistant displays and alarms, supporting the unique relationship between patients and healthcare providers, distinguishing end-user groups, accommodating legacy devices, guaranteeing low latency, allowing for personalization, and ensuring patient privacy. Results are synthesized in design suggestions ranging from personas, scenarios, workflow, and information architecture, to prototyping, testing and iterative development. Finally, future developments in smart healthcare and Human Computer Interaction design for mobile health devices are outlined.

**Keywords**

design methodology, human computer interaction, medical services, mobile computing, user centered design, user interfaces.

# 1. Introduction

The introduction of smart technologies, such as the Internet of Things (IoT) and cloud computing, provides opportunities for advances in healthcare device development [1]. This has the potential for significant improvements in the level of healthcare provided, and in turn to improve healthcare outcomes [2].

However, like all new technologies, their introduction adds a learning burden and cognitive load for both patients and healthcare providers (HCPs). Human computer interaction (HCI) design aims to find solutions to such problems. HCI in mobile health (mHealth) devices is a key process in the interpretation of health data, and dissemination of health services, as it shapes the connection between these devices and both the patients and HCPs [3]. Good HCI design of smart mHealth devices provides future healthcare systems with long-term health monitoring solutions, and lays the foundation for rapid response and interventions in the event of emergencies [4].

## 1.1 Background

Traditionally, research into medical device design was concerned with ergonomics and usability [5], as this was a requirement for regulatory approval [6]. The application of computer science into clinical medicine led to more complicated electronic devices, where HCI design emphasized the interaction between HCPs and medical devices, or between HCPs and computer systems [7].

Nowadays, smart technologies have led to substantial changes in how medical devices are designed and deployed [8]. The evolution of smart mHealth with sensors, big health data, cloud, social networks, health apps, and mobile communication systems, is transforming healthcare services into personalized care delivery [9]. The new medicine model can be defined as 10 P: personalized, perspective, predictive, preventive, precise, participatory, patient-centric, psycho-cognitive, postgenomic, and public [10][11]. A new model of medical service delivery is imminent, and centers care delivery around smart healthcare services and their users [12]. Smart devices, whether used by patients or HCPs, are essential components for the implementation of smart healthcare services and models [13].

The world's largest information technology and internet companies are starting to identify opportunities to invest in the medical field by leveraging their expertise with IoT, cloud computing, and artificial intelligence. The global mHealth market is projected to reach USD 63 billion by 2021 [14]. Global adoption of IoT in healthcare is expected to reach a market of nearly USD 410 billion by 2022 [15]; and IoT in healthcare is expected to grow to a market size of $300B by 2022. HCI design provides interfaces of user control, data visualization, and comprehension in these domains [11][16].

## 1.2 Existing research

There is an increasing amount of research in smart healthcare. mHealth devices have shown great value in managing diabetes, asthma, depression, hearing loss, poor vision, osteoarthritis, anemia, and migraines [17]. Mobile communication and ubiquitous computing technologies allow us to reconsider key healthcare concepts [13]. Mobile devices, combined with a smart healthcare system, can contribute significantly to improving diagnosis, treatment, rehabilitation, and prevention; they can enhance interactions between patients and HCPs, facilitate self-management of chronic diseases, improve translation of health knowledge, support collaborative care, and improve system effectiveness [18]. mHealth applications use mobile and wireless communication technologies to facilitate healthcare services, ensuring enhanced quality, efficiency, flexibility and cost reductions in healthcare delivery [19]. Smart healthcare services require the real-time data sharing, data processing, and data analysis for intelligent decision making; mobile cloud computing plays a vital role to connect services, people, and sensors seamlessly, with major applications lying in critical patient monitoring, telemedicine, and personalized medical services [20].

The use of mobile wireless body area networks can improve healthcare delivery and ensure cost-effective and patient-centered disease management and prevention [21]. Wireless technologies provide greater flexibility and portability compared to traditional solutions [22], and the ability of smart devices to share information more easily provides the possibility of remote diagnosis [23]. However, while smart healthcare is still evolving, there are many significant problems to solve. HCI helps the new technologies be readily available, accessible, adoptable, and acceptable [24]; it helps to ensure that solutions meet user needs and deliver effectiveness and efficiency [25].

### 1.3 Current Issues

As well as focusing on issues of usability, efficiency, effectiveness and acceptance, HCI design research must consider the dependence of human characteristics on technological and social background; the lack of a holistic approach in smart healthcare systems is a factor contributing to the low adoption rate of these technologies [26]. There remains a significant gap between mHealth innovation and user acceptability [9]. Developing effective digital health interventions requires overcoming the challenge of multidisciplinary collaboration between health and HCI experts [27].

HCI and human error issues are critical [28]. Defects in medical devices, including usability problems and poor user interfaces, can jeopardize patients' lives [29] [30]. Analyses of fatality incidents suggests a "lack of insight into design issues is very common" [31]. Industrial design and human factors approaches can reduce adverse outcomes [32], and incorporating HCI principles and guidelines during design can contribute to minimizing defects of medical devices [33].

### 1.4 Aim

This paper contributes an HCI perspective on the design of smart healthcare devices (Figure 1); this is critical to smart systems, because good HCI design enhances both the usability and usefulness, and avoids defects of these devices. The benefit is that appropriate data collection, data management, and use of information within healthcare systems will "determine the system's effectiveness in detecting health problems, defining priorities, identifying innovative solutions and allocating resources to improve health outcomes" [34]. Therefore, we are outlining HCI guidelines for designers of smart mHealth devices. We propose design principles and methods, to complement the user centered design approach. The aim is to promote the development of devices with both higher end-user satisfaction, but also smoother integration into smart healthcare systems, to facilitate efficient connections between users, devices, and smart healthcare services.

## 2. Information architecture for smart healthcare

Smart healthcare combines life sciences and information technology and originated in IBM's Smart Planet strategy [35]; this strategy proposed embedding sensors into a variety of physical objects, to form connections between these objects, and to integrate them through cloud computing [36]. The application of technologies will establish electronic health records that facilitate a) smart tracking of patients, HCPs, and devices, b) patient monitoring, c) patient management, and d) the patient's journey though smart healthcare systems, which will enable a ubiquitous environment with improved healthcare delivery outcomes by connecting social and business data with medical, wellness, and health-related information [2].

While some early smart systems used only three layers (sensing, network, and application layers), it has been shown that in a smart system, separating information transmission and processing layers can facilitate resource sharing and reuse, thus avoiding closed applications [37]. Hence, in this paper, we propose a four-layer information architecture for smart healthcare (Table 1). In this model, each layer depends on its underlying layer.

Placing close-range wireless communication, sensor nodes and gateways into the sensor layer, rather than giving them a fifth layer (or including them in the communication layer), facilitates sensor interconnection, node cooperation, device compatibility, scalability, and data preprocessing; it also aligns with the IoT principles of perceptual connectivity.

**2.1. Sensing layer**

The sensing layer, at the bottom of the architecture, performs the intelligent sensing, monitoring, and collection of health data from medical environments including hospitals, nursing homes, outpatient clinics, medical transports, imaging facilities, and laboratories, as well as from individual patients and families in their homes and communities. The main technology used is IoT, supplemented by biotechnology and nanotechnology for novel sensors [36].

**2.2. Communication layer**

The communication layer is based on the Internet, telecommunication, and broadcasting networks. It combines fiber optic and wireless broadband networks, aiming to provide convenient, high-speed, reliable, large capacity, and comprehensive infrastructure to facilitate data sharing between medical devices, patients, and HCPs [38]. To achieve real-time transmission of medical information, which is required for high quality provision of smart healthcare services, the network could prioritize health data traffic specifically, yet there remains a lot of development work to do and regulatory oversight is still progressing.

**2.3. Data integration layer**

The data integration layer completes the data fusion and data sharing processes, which are a core need for supporting smart healthcare services. It uses a service-oriented architecture, big data and cloud computing technologies to build electronic health records, share health information resources, and augment data with information from other smart systems. The integrated information provides the support platform for the application layer [39].

**2.4. Application layer**

The application layer supports the specific healthcare business needs. The construction of various types of eHealth applications are based on electronic health records, which uses federated data models that are securely linked to facilitate real-time comprehensive information sharing by patients and HCPs, and to make the best use of all levels of available medical resources. It provides users with convenient and efficient links of applications and smart healthcare services, and connects different segments of the healthcare industry.

The users of the application layer include the targets of smart healthcare services (patients and those seeking preventative health services), the healthcare providers (HCPs), management and key decision-makers (public health departments, hospital managers, and drug oversight bodies, such as the Food and Drug Administration or Health Canada), and related institutions (hospitals, nursing homes, community clinics, etc.). Smart healthcare services offer users access to different information channels and interaction modalities, to best fit their needs [12].

There exist some related frameworks: Domingo proposed a hierarchical structure of IoT systems consisting of perception, network, and application layer [40]; the International Telecommunication Union suggested a five-layers IoT architecture with sensing, accessing, networking, middleware, and application [41]. A smart health architecture has been established that consists of Web of Things infrastructure, webRTC compatible browser and a web interface [42]. Firouzi *et al.* presented a smart health architecture consisting of a device layer, fog layer, and cloud layer, and outlined architectural elements including three components of body area sensor network, Internet-connected smart gateways or a local access network, as well as cloud and big data support [16].

## 3. Interaction modes and features of smart mHealth devices

Smart technologies enable application integration and interoperability between different devices and facilities in health systems [43], which have been a significant barrier to smart applications in the past [44]. This in turn will prompt innovation in the field of HCI [45]. Point-of-care testing, wearable sensors, and home-based sensors will allow vast amounts of physiological signals and data to be obtained outside the clinical environment. Cloud platforms enable HCPs to perform remote testing for diagnosis and to gather data for treatment monitoring [46]. Smart healthcare systems extend cross-regional collaboration for diagnosis and treatment, and focus on family and community health management [47]. Healthcare delivery is also transforming from curing illnesses to emphasizing disease prevention and public health improvement [48]; thus a variety of novel healthcare services and models will become possible. These changes create the need for new interaction modes between patients and HCPs [49], with scenarios and interaction processes different from traditional electronic clinical tools, such as Virtual Health visits. Smart mHealth devices are able to know when and how to offer services according to personalized requirements in ubiquitous scenarios depending on different user needs [50].

### 3.1 User groups

For patients, smart mHealth devices aim to: a) collect physiological data using sensors connected through body area networks; b) transmit these data to those storage and processing locations with the lowest latency that can be achieved cost-effectively; c) make physiological monitoring information available anytime and anywhere; d) provide smart management of personal health information; and e) build personal health information assessment and health optimization feedback systems [39].

For HCPs, these devices have the potential to: a) view patient information in real time; b) analyze patient data remotely, or when physically present; c) optimize emergency services and local first aid delivery; and d) ensure timely response with optimized allocation of cross-regional medical resources.

The challenge is how to meet users' needs, choose appropriate interaction methods, provide rich information, and integrate smart features simultaneously. The vision of smart healthcare, with its demands for widely-available information, architectural frameworks, delivery modes, and technological opportunities and constraints, must be balanced in smart mHealth devices, along with the inherent key healthcare properties of privacy, currency, urgency, trust, and safety. Smart mHealth devices have different features from other intelligent devices and thus are potentially more demanding.

### 3.2 Key features of smart mHealth devices

One important feature is that, besides the usual interaction approaches, which focus on users' visual, auditory, and tactile channels, other interaction approaches may need to be added according to the sensors and the user's medical condition. Some diseases or medical treatments may impact the user's vision, hearing, tactile sensitivity, and cognitive abilities. On the other hand, some biomedical sensors can capture physiological signs data without the users' attention, passively in the background. Thus new interaction approaches and features are emerging.

For example, a wearable patch with microelectronics technology and polymeric substrates, applied as a disposable tattoo to the users' skin, can capture respiration rate, skin hydration, muscle movement, heart rate, and brain activity for a week [51]. A smart pill with an edible microchip, activated by stomach acid after swallowing, can detect medication compliance, sleep patterns, and physical activity levels [52]. These devices can provide HCPs with reference data on a patient's condition; and provide patients with personalized medical programs and medication reminders.

Scalability is another important feature. The overall structure is supposed to adopt an open design, using identifier fields to distinguish different types of devices, while complying to medical ontologies. In addition to common fields, provided by most devices, devices can also have their own custom extensible formats, for which they need to providing data dictionaries. Correspondingly, the user interfaces need to facilitate expanding functionality, while maintaining consistency. This allows users to add plugins and customize devices to their unique needs; and make updates, modifications, improvements of the device easier, enabling rapid system growth [18]. Connectivity issues and different network interfaces for communication with intelligent auxiliary hardware components must also be considered.

Various constraints will impose challenges, including limited device size, battery life, memory, storage space and computational processing power. The support of cloud computing resources may mitigate some of these issues. Similarly, the device display may be restricted by limited screen dimensions, resolution, and color performance, which pose challenges to displaying a large amount of information. This is more challenging when there is more content added to already-small screens, because the display and control panels of mHealth devices may be combined into a single entity, in order to reduce the device size and improve portability. In some cases, sensors can be separated, performing only the sensing and data acquisition, with limited computational processing; while other components are used for analysis, control, display, and storage. This is different from traditional devices that often have all the components in a single unit.

## 4. HCI design principles for smart mHealth devices

The HCI design for smart mHealth devices is complicated by the need for a large variety of interaction modes to support different user roles, tasks and workflows. Features of a smart healthcare environment and system functions for multi-threaded, dynamic physiological data collection, require secure, fast and real-time transmission, reception, storage, and display that are compliant with healthcare frameworks. Designers need to strike a balance between the goal of providing a smooth-running data acquisition platform, and offering mobile client-side services. Smart mHealth devices thus require unique HCI design principles.

### 4.1. Safety vs. Ease-of-use

Safety considerations are important factors for smart mHealth devices. Usability refers to effectiveness, efficiency, and satisfaction [53], improving ease-of-use of the design [54]. Heuristic evaluation standards include efficiency of use [55]. However, for medical devices, this efficiency or ease-of-use may conflict with the device's safety. Ease-of-use encourages the use of cues to minimize cognitive load, to act quickly or even automatically, which means "response chaining" and "muscle memory" may be used without requiring a cautious check of the task. Such behavior might not be desirable for safety, which is critical to medical devices. For example, the Therac-25 radiation therapy machine exposed patients to hundreds of times excessive doses of radiation, resulting in death or serious injury; one of the contributing causes was that the original redundant design of entering control parameters twice could be circumvented in the latter version by users trying to speed up their tasks [56].

The critical nature of healthcare can magnify the potentially negative outcomes of safety and security risks, which can be mitigated by HCI design. The use of data validation and automatic checks may improve safety whenever a device receives and sends data, and prompt the users to rethink their action. Restricting the use of unsecured wireless networks or cloud storage may protect personal health record data from the threat of cyber-theft. Requiring passwords before operating the device may keep unauthorized individuals away from critical control settings, though it might also delay use and cause login-fatigue. Software routines that detect possible dangerous settings or input data may prevent non-reversible mistakes, though they might also contribute to alarm fatigue. Additional notification and authorization protocols may support required actions and workflow, e.g. a nurse may need authorization from a physician for a given treatment; or, in some cases, actions may need to be double-checked and authorized by a colleague. Finally, a user interface, which imposes a strong sense of safety, can enhance the confidence of HCPs in the device, and potentially reduce anxiety of patients using it.

**4.2. Error-resistant displays and alarms**

Use errors result in outcomes that are not intended by the manufacturer or expected by the user. If safe and effective use is not achieved, use error occurs [57]. Here, use error refers to user interface designs that cause users to make errors of commission or omission [58].

There are many instances of devices recalled by the U.S. Food and Drug Administration (FDA) because the visual display of user interfaces led to use errors [58]. For example, the graphics of a software system made doctors confuse the patient's left and right hemispheres of the brain; this may lead to wrong-side brain surgery. Similarly, the ability to change the units displayed on a glucose meter between mg/dL and mmol/L, which differ by a factor of 18, may lead to misinterpretation of the measured value and incorrect therapeutic action. Finally, the numbers displayed on a glucose meter made patients misunderstand their blood glucose levels because the "2.2" was too similar to "22"; this may lead to much higher insulin dose administration, causing severe hypoglycemia, diabetic coma or death [59].

Ineffective audible and visual alarms of medical devices may lead to HCPs potentially ignoring critical situations and not taking prompt action. In fact, false alarms are considered to be the number one health technology hazard [60], as they interrupt clinicians during critical tasks [61], cause distractions and alarm fatigue [62], and create noises stressing patients and HCPs alike [63].

The consequences of design flaws in medical devices are potentially disastrous. A critical component is the user interface, through which users understand the inner workings of the system, and its state. Designers can predict, prevent, and minimize use errors by iterating usability tests and analyzing contextual risks; this includes the potential use errors during actual use, the frequency of the potential errors occurring, and the severity of the resulting harms. It is possible to predict use errors by simulating the users' environment, exploring the interaction in scenarios (including abnormal and anomalous conditions), and considering all users, devices and systems in the environment. It is also important to be aware that users may be distracted, ignore prompts, and forget important information or experience fatigue under difficult physical and psychological conditions.

It is useful to test various modes of alerting separately, such as visual only, sound only, and special device behavior when encountering a low battery state. Designers are supposed to prevent the disabling of life-critical alarms, or at least ensure the user is aware of the disabling state. Critical information should be legible and readable in the intended use environment. The design of buttons, fonts, typesetting, and layouts should be conducive to good judgment and allow for timely intervention with few errors. When errors occur, a user interface should ensure that the inner state of the system is visible, that users are given enough useful information to deal with the problem, and that files are saved automatically so recovery can be performed without data loss.

Norman divided errors by the level of users' intention to act. Inappropriate intentions cause mistakes while unintended actions cause slips [64]. Sternberg found that slips occur when users deviate from routines, automatic processes override them, or they are interrupted [65]. Many disastrous cases fall into the category of slips rather than mistakes. HCI designers should not simply think of the difference between errors and correct behavior, but consider the entire interaction of users, devices, and systems, as one error might cause errors by the other entities involved [66]. There are some points from Nielsen's ten heuristics [67], Norman's seven principles [49], and Shneiderman's eight rules [68], which are useful for creating error-resistant design. For example, designers can help smart mHealth devices minimize potential errors by alarming or flagging actions that may result in errors and help users recognize, diagnose and recover from errors by offering a digital equivalent of taking a step backward, or a clearly marked emergency exit, or by informing the user of the operational status and suggesting solutions in a timely, highly visible and comprehensible way. The Therac-25 is also an example of poor error-resistant design, showing useless error messages [56].

**4.3. The healthcare provider-patient relationship**

Multiple relationships that exist between patients, HCPs, and devices, can impact the user experience. Thus, special design strategies are used to support these relationships.

The first group of users for smart mHealth devices is HCPs, who have medical knowledge and are familiar with the operation of medical devices. Within this group, there are different levels of experience, expertise and training, which result in different modeling of roles and might require different user interfaces within the same device. These users have strong requirements for information clarity and organization so that they can comprehend the information they most need quickly in order to make rapid decisions. Controllability and advance warning are important requirements so that HCPs can respond quickly to changes in patient information. To facilitate these requirements, user interfaces should be easy to control, have clear and accurate information display, be supplemented with appropriate and intelligent reminders, conform with HCPs' mental models, and enable ethical use with accountability, security, and protection of patient privacy [69].

The second group of users are patients, who have different characteristics, knowledge, and different health conditions imposing potential constraints on their capabilities. For them, a good design is simple, secure, easy to read and operate, respects the user's abilities and conditions, does not add additional stress, and emphasizes universal design, advocating inherently free access for people with disabilities [70].

The third group of users includes non-professional caregivers, e.g. patients' family members and domestic helpers. Here, the design needs to ensure that it is easy to follow HCPs' guidance because their interaction with the device will indirectly affect patients' health outcomes, or directly determine the healthcare intervention as the patient loses the ability to complete tasks independently due to changes in physical and mental status.

For example, Propeller Health's digital device and application tracks symptoms, medication compliance, and drug use patterns for patients with asthma or chronic obstructive pulmonary disease (COPD). The sensors of the inhalers collect data and document synchronized information about the patient's condition, time, location, and use of inhaled medications. Patients manage their own information and monitor their own treatments. Collected information is used to generate reports, which are shared with HCPs who then manage changes in the patients' treatment [71].

Another example, Bodytel, provides diagnostic services to patients with chronic diseases; the suite of monitoring devices includes a Bluetooth-enabled blood glucose meter, blood pressure cuff and a scale. There is a mobile app displaying data on the patient's smartphone and an online portal using a medical data cloud for connectivity. Patients send their data to their HCPs, who configure individualized alert thresholds on the patients' systems according to their baseline data and treatment goals, so that warnings can facilitate timely intervention or rescue [72].

**4.4. Distinguish end-users**

As in general software development, HCI designers must balance potential benefits with feasibility; and identify and resolve the conflicting requirements between different stakeholders. At the same time, they should not mistake customers for end-users. End-users typically do not have the technical knowledge to support, maintain, or design the device, and are oftentimes not the ones who purchase devices either [73]. Different end-users have different needs and attitudes towards new devices. Investigating end-users and the environment where the device will be used is an important aspect to ensure device usability.

Early focus on end-user needs, and the tasks they use the device for [57], helps to avoid serious operational problems, and facilitates making the device easy to use, easy to learn, and easy to remember. The FDA expects summative usability testing to involve at least 15 representatives from each distinct end-user group [73].

**4.5. Legacy support**

Unlike other smart devices pursuing innovation, medical device designers must be cautious in implementing changes, because of long-term stability and durability of medical devices. The medical product life cycle is long and the introduction to market is slow. It has been estimated that the time from concept to market is 3 to 7 years [74], while most mobile phones have a 9 to 18 month market life cycle [75]. Through long-term use, users form cognitive habits and develop expectations of typical interaction methods. HCI designers can leverage users' old workflows, refer to legacy systems' methods and standards, and seamlessly integrate them into the new environment and system. Leveraging existing features will help to optimize designs, minimizing the user's learning curve, re-using existing mental models, and maintaining a stable transition during the introduction of new medical devices.

For example, the Withings' Wireless Blood Pressure Monitor has the appearance of a traditional non-invasive blood pressure cuff. It leverages the existing cognitive habits, facilitating immediate and efficient operation of the new device; it is inflated automatically like the traditional device, but the display is moved to the screen of a smartphone and there are no rubber tubes involved. It measures users' heart rate, systolic and diastolic blood pressures, and pulse pressure variability, and visualizes the data in a way that allows patients to compare results and consult with their HCPs [76].

**4.6. Timely response**

Time plays a key role in ensuring effective health care delivery. To address different aspects of timeliness and currency, each layer has additional constraints. In particular, the sensor layer must minimize the latency between a value being measured, and its availability to the rest of the system. In some health contexts, like emergency rooms, the actionability of a result depends upon the data being communicated within a period of real-time constraints, placing demands on the communication layer. When aggregating data from different sources, the data integration layer must ensure data is relevant and current, and the application layer needs to make the user aware when timeliness and currency constraints are not being met, such as including timestamps and data frequency indicators.

Smart mHealth devices must respond swiftly to changes in the users' status, quickly switch to the corresponding program mode, and be easy to monitor, view, check, and control. Providing timely information and efficient response can reduce costs, save time, improve care, and also improves patients' confidence in the technology. An efficient user interface design facilitates rapid decision-making, task-completing and mode-switching, which is especially important in high-stakes triaging situations such as in emergencies. This requires straightforward control features and clear logical displays, in which data can be rapidly interpreted and compared. For example, the Lifestone captures, charts, and synchronizes data of body temperature, heart rate, respiratory rate, stress indices, blood pressure, oxygen saturation, and heart and lung sounds through multiple sensors. It provides patients with their real-time physical condition, and HCPs with comprehensive information to assist diagnoses and treatment planning [77].

### 4.7. Personalization and privacy

Personalization is an important requirement in smart mHealth devices. The device may save and recognize the users' identities and physical characteristics, mine the users' data, learn the users' behaviors, habits and preferences, and customize its user interfaces according to the users' knowledge, capabilities, and skills. Taking all these human factors into account can facilitate good HCI design and provide a better user experience [78].

This personalization is also manifested in individualized designs aimed at patients with different medical conditions, disease prevention and treatment goals. For example, the Valedo Back Therapy Kit, used to treat and manage non-specific back pain, general muscular deficits of the trunk, chronic kidney problems, and spinal cord injury [79], includes two sensors, a gaming platform, a cloud infrastructure, and an intelligent connection module. Patients attach the sensors to their back and chest and the system modifies the game strategy to adapt to their individual situation, allows them to exercise with the motivation of a game, while obtaining feedback on their performance. Meanwhile the exercise data are sent for their HCPs to analyze; individual treatment strategies are adjusted accordingly [80].

Finally, HCI designers need to consider the privacy of personal information, prevent data leakage into the public environment and reduce the risk of data theft through data encryption, secure transport channels, user authentication, and data access authorizations [81].

## 5. HCI design methods for smart mHealth devices

HCI design includes specifying context and requirements, defining personas and scenarios, designing workflows and user interfaces, as well as prototyping, evaluating, and iterating designs. User centered design (UCD) or user-driven development (UDD) should be adopted throughout the problem-solving process, which keeps the user at the focal point during the whole product cycle [82]. It refers to human-centered design (HCD), which is defined by ISO 9241-210:2010 as an approach that focuses specifically on making interactive systems usable [83]. It uses an iterative process emphasizing usability goals, user characteristics, tasks, workflows, and environments; and follows a series of techniques for analysis, design, evaluation, implementation and deployment phases [82]. The typical steps include specification of the context of use and requirements, creation of design solutions and an evaluation [84]. Activities (see also Figure 2) include:

### 5.1 Plan UCD in system strategies

Investigate and analyze the users, environment, systems and devices, based on the smart healthcare architecture. Select UCD methods, ensure a UCD approach, plan and manage UCD activities [85].

### 5.2 Specify and identify

Specify user requirements. Analyze stakeholders, and refer to users in the application layer when targeting smart healthcare services. Specify stakeholders' requirements, end users' needs, and the relationship of device users and other users in the application layer. Extract requirements from observations, interviews, research reports, past case reports, and relevant legacy user interfaces. Detect and resolve potential conflicts between the requirements. User requirements are noted and refined through ethnographic study, contextual inquiry, usability testing, and other methods [86]. Apply those methods to analyze patients and HCPs, including their characteristics, behaviors, habits, skills, attitudes, goals, etc., to obtain a deeper understanding of their requirements, and define the personas accordingly [66]. At the same time, pay attention to the features of the interactive relationships between patients and systems, HCPs and systems, and patients and HCPs.

Specify context of use. Identify physical, technical, and organizational environments. Define scenarios, contexts of use, and tasks of logical activities within the smart healthcare application layer. Multiple scenarios and tasks are to be included for each user role. Analyze events where users participate across a wide range of smart healthcare processes, and acknowledge that there are different interaction methods for different user groups in different scenarios. Understand differences between emergency and routine use, and how to connect different scenarios to achieve ubiquitous services. Describe features in ubiquitous environments. Develop use cases for commonly encountered interactions [87].

## 5.3 Produce design solutions

Conduct content, task, and workflow modeling [88]. Design the information architecture, combining results from the previous steps and the resources of smart healthcare architecture. Include the data acquired by the device and related resources acquired from the data integration layer and sensor layer. Ideation occurs between analysis and design: designers think broadly, draft as many ideas as possible, explore those ideas, and decide which are desirable for users, viable for the system and business, and technically feasible; designers then think narrowly, selecting the most valuable solutions and refining their ideas. The product positioning and concept design follows full analysis of sensors and mobile devices to be used, i.e., sensors and local communication methods, the functional requirements, interaction modes and features, technology pitfalls and limitations in the sensing and communication layer, as well as lessons learned from similar devices by competitors and related items. Consider the relationship of the device with other devices, applications, systems and services in the sensor layer and application layer. Complete the device's overall function design and divide it into specific functional modules. Ensure the modules facilitate triaging and integrating, and meet requirements for real-time physiological signal acquisition, data processing, data transmission, data storage, and dynamic information display within the system.

## 5.4 Produce prototypes

Design and prototype user interfaces. User experience has at least three aspects: a) content, b) behavior, and c) form [89]. The corresponding aspects of interface prototyping are: a) content design, b) interaction design, and c) visual design. The prototypes can test whether the user interface fully meets the users' needs. Low-fidelity prototypes are tentative and include summary content and stand-in images, but no interactivity. They are valuable for refining user requirements. Medium-fidelity prototypes have more detail, but still have limited functionality. High-fidelity prototypes, include interactivity, complete visuals and full content, and are used to test workflows, interactive elements, and graphical components such as visual information hierarchy, affordance, and legibility.

## 5.5 Evaluate design solutions

Redesign and iterative design are emphasized to improve design solutions in every step when moving from a low-fidelity to a high-fidelity prototype. The HCI design loop starts with planning, analyzing, specifying, ideation, and modeling; next follows an iterative cycle of prototyping and evaluating; returning to the design stage is common; even returning to the specification stage may be required. Through repetitive testing, participatory design, and refinement, user interface problems are addressed before they become too expensive to resolve or before they cause serious problems. The UCD method encourages constant user feedback. This approach helps minimize risk, and ensures that user needs are met.

## 5.6 Develop and implement

Make the device perform well in the application layer, be compatible and connective with other sensors and devices in the sensor layer, and communicate well with local endpoints and the cloud through appropriate networks in the data integration layer. Usually, the first few editions will not achieve acceptable performance for deployment, even though the usability of a high-fidelity prototype may have been tested through several rounds and be regarded as perfect. The users' real requirements become much clearer during actual use of the device. Serious problems, risks, inconsistencies and misunderstandings are identified during the product's early lifecycle. Thus, designers need to collect data after each version is deployed, detect the effectiveness, efficiency, and user satisfaction, identify problems, leverage lessons learned, modify design solutions, and improve the device continuously.

**5.7 Evaluate in use**

Test usability, evaluate user experience, iterate design and iterate development, while emphasizing the smart healthcare context. The goal is to optimize usability and the user experience, and to achieve a satisfactory device that meets users' requirements, and fully provides the targeted healthcare service.

For healthcare devices, testing is often one of the most challenging parts to accomplish with patients, HCPs and administrators. Administrators play an important role, and this poses challenges for new healthcare service provisioning in the US, so the requirements and regulations should be considered early. The ability of patients, their willingness to participate in testing, and their performance during testing varies. They may be impacted by their physical, mental, sensory and cognitive status, comorbidities and medical conditions; additionally, privacy and health considerations might limit their enthusiasm to participate.

HCPs' time is limited, so testing is often conducted in clinical environments with distracting noises, alarms, clutter, workload, urgent events, and colleagues. In order to obtain useful feedback quickly, evaluations generally rely on field interviews in rushed conversations using medical vocabulary. Interviewers often lack medical expertise or extensive HCI training. Though it is an effective method to obtain information of user expectations, it may not address the usability issue of efficiency [90]. Personality and style of the interviewer may affect the response; respondents may seek to please the interviewer [91]; and the resulting data may be misinterpreted [92]. Similarly, testing needs to differentiate between testing of a device's usability, and testing of users' learning abilities. Thus, testing can be improved by collaboration between designers, developers, and HCPs.

Appropriate testing methods will save time and resources. Some usability inspection methods and empirical user testing methods are adopted. Firstly, tests by expert evaluators, using the cognitive walkthrough method to work through a series of tasks step-by-step on the device, along with thought-provoking questions from the perspective of the user [93]; and secondly, using the heuristic evaluation method by comparing the device to a set of heuristics principles to identify and resolve problems at an early stage [94]. After these, test the device with patients and HCPs, mixing both field tests and lab-based tests, where audio recording, video recording, and eye tracking are possible. Finally, test results must be interpreted cautiously, considering the factors of users, testers and environments.

During the entire HCI design process (Figure 2), the UCD approach is emphasized at every step from analysis to evaluation. Some stages overlap and they are not considered separately during execution. For example, analyzing, specifying, defining, and modeling are all included in the user interface design and interaction design. Persona, scenario, and use-case are all analysis tools in the UCD approach. They are essentially a prerequisite, and thus are required before prototyping begins. The visual design is included in user interface and interaction design. Finally, designers participate in the evaluations, adopting a bridging role between the mental models of users and developers.

The purpose of the UCD approach is threefold: to create devices that are useful, usable, and used. 'Useful' is having a practical or beneficial use; allowing a user to accomplish a task. 'Usable' is being capable of being used, allowing a user to use the device in a pleasant, simple, learnable, intuitive, convenient and effective manner; it refers to the usability, user-friendliness and accessibility [95]. 'Used' is the ultimate aim because a design must be used by users otherwise it is considered a failure. If a device is usable, it increases the chances of being used, though there are many other factors that impact device adoption in the mHealth market.

## 6. Discussion of UCD approach in HCI practice for smart mHealth devices

This paper contributes to the work of HCI designers and developers who aim to create better smart mHealth devices by emphasizing UCD throughout the HCI process.

### 6.1. Advantage of using a UCD approach

The UCD approach has shown substantial return on the investment in time and resources required in many successful examples. It is reported that 15% of information technology (IT) projects are abandoned, and more than 50% of a programmer's time is spent on reworking the design and implementation, which results in billions of dollars spent every year on preventable failures [96]. This problem is likely getting worse, as IT is becoming ubiquitous. Primary reasons for failure include unrealistic or poorly articulated project goals, poorly defined system requirements, and ineffective communication between customers, developers, and users. The cost of fixing an error when it is discovered in the field is 100 times higher than when mitigated during the development stage [96]. HCI practice can help avoid costly large-scale rework by defining requirements and modeling reality from the beginning, identifying problems at an early stage, finding solutions during each cycle's evaluation, and making iterative improvements with users' participative communication. Large IT failures of smart healthcare, resulting from HCI omission, are more than an expensive inconvenience – they can put lives at risk. Human Factors International (HFI) recommends following the "10% rule", whereby 10% of IT staff should be user experience (UX) professionals, and 10% of the budget dedicated to UX [97]. The user-centered or human-centered methodology's benefit is clear: more usable and receptive designs, improved usability, less failures, and faster learning of the new device. It considerably reduces back-end costs though adding a few up-front costs.

The UCD approach is particularly applicable for mHealth devices. According to ISO 9241-210:2010(E), this approach "enhances effectiveness and efficiency, improves human well-being, user satisfaction, accessibility and sustainability; and counteracts possible adverse effects of use on human health, safety and performance" [83]. Features of the UCD approach with human factors design and evaluation are mandated by the FDA for medical technologies [98]; usability test is recommended by the Agency for Healthcare Research and Quality to ensure safety and effectiveness [99]. mHealth offers opportunities to improve healthcare delivery and outcomes by meeting specific users' requirements at the right times and right places. The most significant factor is that active engagement and self-management allows timely interventions based on behaviors and responses of patients and HCPs under certain conditions. UCD can enhance the intervention effectiveness of a device by engaging users in the development process, including the investigation of user requirements and user participation in evaluation. Finally, the World Health Organization recommends user evaluation in mHealth projects [100].

### 6.2. Disadvantages of a designer-centered, technology-centered, or system-centered design approach

Unlike the UCD approach, other methods are driven by the interpretation of designers, the capabilities of technologies, or the implementation of system thinking. The UCD is the only approach that keeps users at the focus of the HCI process, thus facilitating the creation of a simple and straightforward device for users.

The designer-centered approach is not well suited to smart mHealth devices, though it can create innovative devices with the latest technologies. Due to the diversity of users, including patients, their families, and HCPs, there is not a specific user population with unique requirements for which the designer-centered approach works well. In addition, the designer-centered approach increases the risk of developing devices that are hard to use or are not compatible with other devices or smart healthcare systems.

There is a similar problem with technology-centered design, which emphasizes the application of technology, which may not be useful or desirable. In both approaches, users' needs are rated lower than the needs of creativity and expression of design teams or technology teams.

Finally, the system-centered design focuses on the functionality of the overall system, implementing system thinking in design practice to address problems of complex systems. Without necessarily thinking about users, the information and structure of user interfaces might cause confusion because the users' knowledge is assumed incorrectly. Also, the users' learning abilities are often overestimated in this approach.

### 6.3. Disadvantages of using a UCD design approach

The UCD approach does have some weaknesses. It might lead to less cohesive and more complex designs that do not consider activities and their sequence. It is not good for situations in which significant organizational constraints are imposed, as is often the case in healthcare. Next, focusing on certain individuals might ignore others' needs. Finally, user needs change with increased proficiency, which can be hard to anticipate during the UCD. Don Norman, who coined the term UCD, said "listen to customers, but do not always do what they say" [101] and Jakob Nielsen said "users do not know what they want" [102], so pay attention to these potential pitfalls.

### 6.4. Advantage of integrating a UCD approach in smart healthcare system architecture

We suggest keeping the end-users as the primary design stakeholder and focusing on what the users' really need, while at the same time considering system thinking, where users are embedded in the architecture of the system as important participatory components [103]. Designers not only incorporate user characteristics, behaviors, motivations, and activities, but also incorporate the interconnection, interaction, consequences, and context in the system architecture. Currently, many mHealth interventions are based on existing healthcare systems, which are developing towards smarter healthcare systems. Therefore, we aimed to describe general processes and principles, which can be applied broadly in a variety of different cases of smart mHealth device design.

Finally, the requirements of users, businesses, function, service, aesthetics, and implementation often conflict with each other. Thus, a balance between them needs to be struck; the balanced solution can only be found if designers are aware of the conflicting requirements, and maintain a clear conceptual model.

Using the UCD methodology with smart healthcare system architecture leverages benefits of ensuring users can understand and use the information provided by a complex system. The users' feedback should be acted upon, and their needs are identified, while organizational and system goals are still met. Using this method can also increase the operational efficiencies of the whole system, improve analysis of performance and feedback measurements, help reduce system complexity, deliver services to patients more quickly, simplify the health-seeking process, and reduce reliance on limited healthcare resources.

## 7. Recommendations and future work

Smart healthcare is an emerging field. HCI study and application in this area not only benefits the development of smart healthcare, but also expands the theory and practice of HCI as an interdisciplinary subject [89]. This paper provides insights and strategies for HCI practitioners when developing smart mHealth devices, or when IoT and cloud computing technologies are added to existing medical systems. Theoretical research can provide useful guidance, yet as the field is in an exploratory stage, there are few cases available to learn from. Enriching research in practical design and development is an important future direction for HCI researchers and practitioners.

Smart technologies have enabled the rapid development of smart mHealth devices in the recent past, and increased demand for upgrading existing medical devices. Future smart mHealth devices are likely to be completely different from today's hardware and software systems, and the way people perceive and interact with them will be different as well. Smart systems will likely fully integrate into everyday objects or buildings. Autonomous behavior such as the automatic provision of healthcare services using sensor data and self-learning clinical decision support systems will become characteristics of smart healthcare services. HCI addresses this issue, as a loss of control by users has the potential to undermine the integrity of the whole system, and both patients and HCPs will subvert the system with workarounds. This may be particularly true in healthcare and poses a major challenge for HCI, more so than for other aspects of IoT and cloud computing.

Finally, current mHealth devices will develop towards a more intelligent, more convenient, and more natural smart healthcare technology. The HCI methodology must keep pace with these developments.

## 8. Conclusion

Including an HCI approach in the development of smart healthcare devices is critical, because good HCI design enhances both the usability and usefulness of these devices. HCI design in smart healthcare should be delivered with a user centered design approach, and is most effectively supported by the system thinking of a four-layer information architecture, consisting of application, data integration, communication and sensing layers. Specific considerations for HCI in healthcare include a) balancing safety with ease-of-use, b) the need for error-resistant displays and alarms, c) providing timely response with minimized latency, d) satisfying the varying requirements of different user groups, while e) respecting the healthcare provider-patient relationship, as well as allowing for f) personalization and security, and g) integration with legacy workflows and systems.

# Tables and Figure Legends

**Table 1.** Four-layer information architecture of smart healthcare

| Layer | Application/Example |
|---|---|
| Application Layer | Applications: regional healthcare platforms/dashboards, epidemic/outbreak detection systems, community healthcare services, personal health records, home health monitoring systems, smart hospital systems<br>Delivery mode: devices, applications, dashboards, websites, electronic medical record |
| Data Integration Layer | Medical information resources, cloud platforms/distributed data storage, data sharing, data processing, data fusion, data extraction, statistics/predictive modeling, data analysis applications |
| Communication Layer | Internet, telecommunications networks, broadcasting networks, wireless networks |
| Sensing Layer | Sensors: CCD/CMOS imaging sensor, RFID, SCADA, GPS, medical sensors<br>Local communication: WBAN, NFC, wireless network, Bluetooth, Zigbee |

CCD: charge-coupled device, CMOS: complementary metal-oxide semiconductor, RFID: radio frequency identification devices, SCADA: supervisory control and data acquisition, GPS: Global Positioning System, WBAN: wireless body area network, NFC: near-field communication

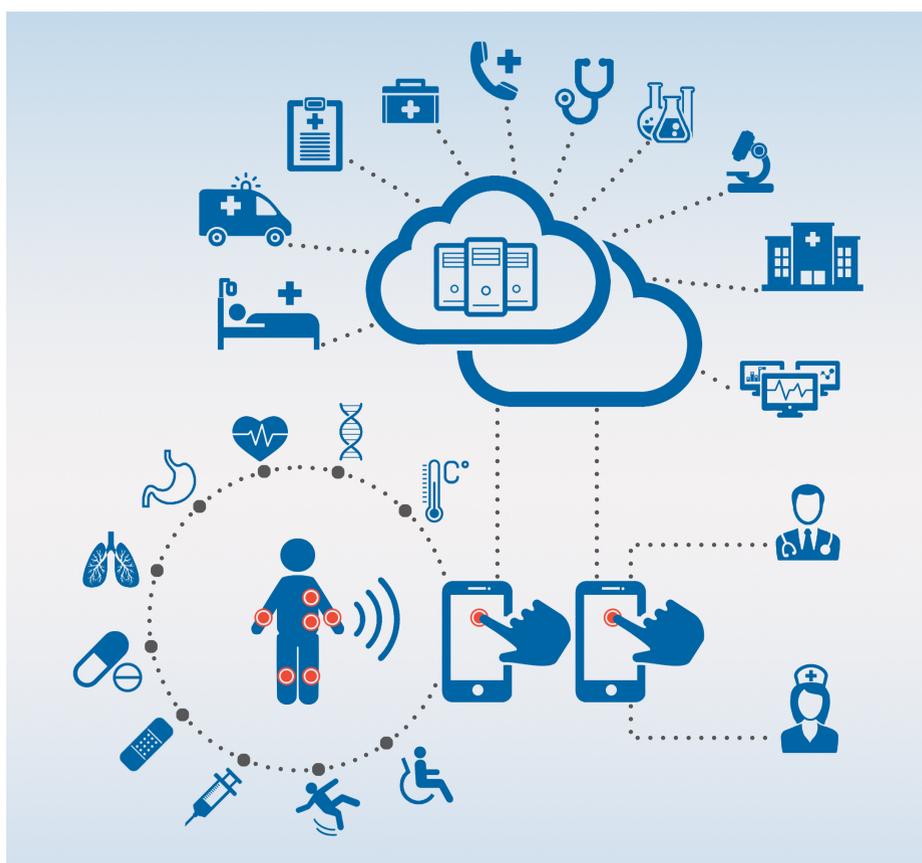

**Figure 1**. Human computer interaction in smart healthcare mobile devices, with the concept of IoT-based and cloud-based healthcare services and applications.

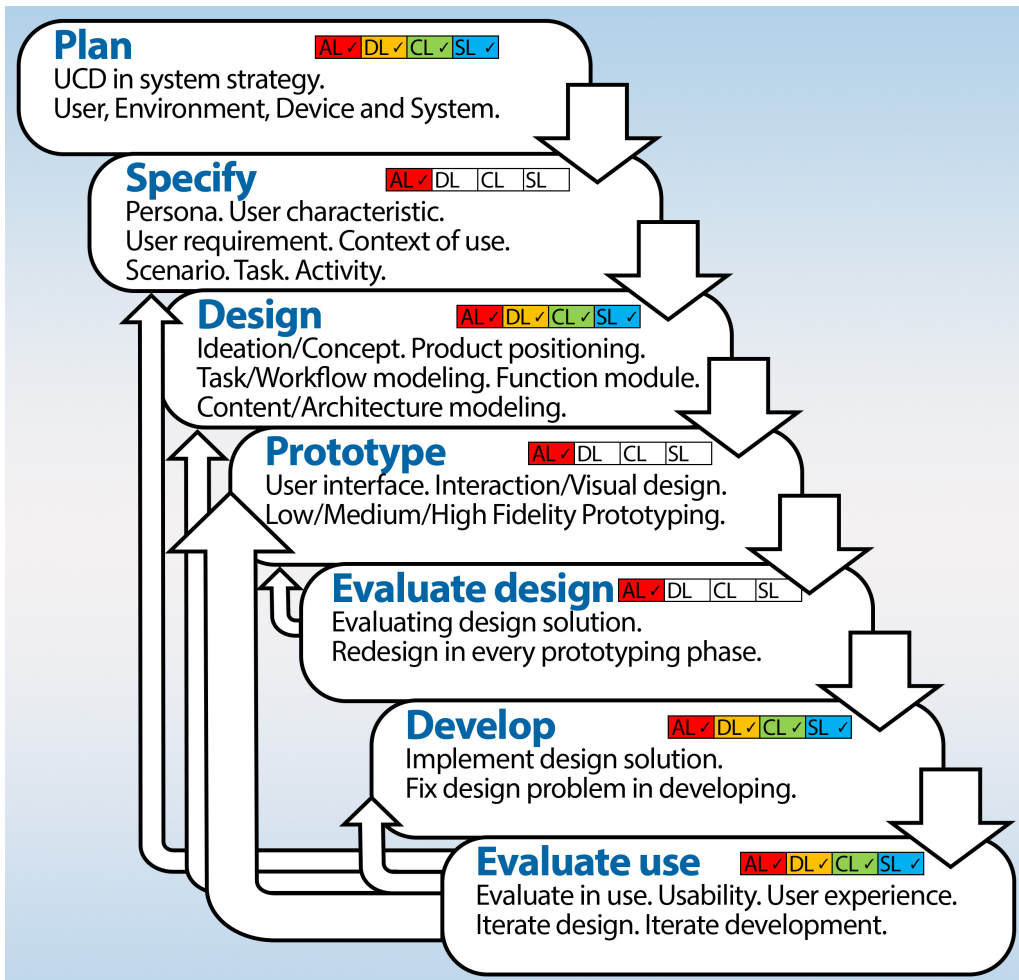

**Figure 2.** Process of HCI design for smart mHealth devices and the key smart healthcare architecture layers involved at each stage (AL: Application Layer, DL: Data Integration Layer, CL: Communication Layer, SL: Sensing Layer).